\documentstyle[aps]{revtex}
\input{epsf.tex}
\begin{document}
\draft
\title{Self-consistent theory for molecular instabilities in a normal degenerate Fermi gas in the
BEC-BCS crossover}
\author{R. Combescot$^{\,a}$, X. Leyronas$^{\,a}$ and M.Yu. Kagan$^{\,b}$}
\address{(a) Laboratoire de Physique Statistique,
 Ecole Normale Sup\'erieure*,
24 rue Lhomond, 75231 Paris Cedex 05, France}
\address{(b) P.L. Kapitza Institute
    for Physical Problems, Kosygin street 2, Moscow, Russia, 119334}
\date{Received \today}
\maketitle

\begin{abstract}
We investigate within a self-consistent theory the molecular instabilities arising in the normal state of a homogeneous degenerate Fermi gas, covering the whole BEC-BCS crossover. These are the standard instability for molecular formation, the BCS instability which corresponds to the formation of Cooper pairs and the related Bose-Einstein instability. These instabilities manifest themselves in the properties of the particle-particle vertex, which we calculate in a ladder approximation.  To find the critical temperatures corresponding to these various instabilities, we handle the properties of the interacting Fermi gas on the same footing as the instabilities by making use of the same vertex. This approximate treatment is shown to be quite satisfactory in a number of limiting situations where it agrees with known exact results. The results for the BCS critical temperature and for the BE condensation are found to be in fair agreement with earlier results. The threshold for formation of molecules at rest undergoes a sizeable shift toward the BEC side, due to quantum effects arising from the presence of the degenerate Fermi gas. This should make its experimental observation fairly easy. This shift remains important at least up to temperatures comparable to the Fermi energy of the gas.
\end{abstract}

\section{INTRODUCTION}

Experimental progress in the field of the BEC-BCS crossover in ultracold fermionic gases has been going on recently at a very fast pace \cite{levico}. In particular after the observation of the Bose-Einstein condensation of diatomic molecules \cite{becmol} both in $^6$Li and in $^{40}$K, evidence for superfluidity on the BCS side of the crossover has been provided by the study of collective modes by Bartenstein \emph{et al} \cite{grim}, which shows a strong attenuation peak, whose likely interpretation is pair breaking in the vicinity of the BCS transition. Further evidence have come from "projection" technique experiments \cite{proj} where a fast sweep from the BCS to BEC side allows to infer the existence of superfluidity on the BCS side. Very recently the observation of vortices in these Fermi gases on the BCS side \cite{vortex} has given a much clearer evidence for superfluidity.

One major interest of the study of the BEC-BCS crossover is to obtain a clear and precise picture of the way in which Cooper pairs go progressively into diatomic molecules when the strength of the attractive interaction between fermionic atoms is progressively increased, and in particular to explore if the BCS formalism gives a proper description of this evolution. The occurence of this crossover within the BCS formalism and its interest has been put forward by the works of Leggett \cite{legg}, and Nozi\`eres and Schmitt-Rink \cite{nsr}, and further investigated by Sa de Melo \emph{et al}\cite{sdm}, although the fact that the BCS formalism gives also a correct description of the condensation of  molecules in the strong coupling limit of dilute molecules was known much earlier \cite{kk,eagles}. 

Apart from its fundamental physical interest this question is also highly relevant for high $T_c$ superconductors since in this case Cooper pairs are known to be quite small and much closer to molecules than in standard superconductors. The situation in high $T_c$ superconductors would accordingly be close to a Bose condensation of molecules, which would provide a natural explanation for the high value of the critical temperature. The natural simple theoretical framework for this solid state problem is the attractive Hubbard model and the self-consistent $T$-matrix approximation \cite{max1,max2} used to obtain a qualitative analytical understanding of the Hubbard model corresponds exactly to the framework we will use in the present paper to handle the case of fermionic atomic gases. 

In practice the BEC-BCS crossover is realized experimentally by going through a Feshbach resonance, by varying a static and homogeneous magnetic field applied to the atomic gas. Due to the very low temperature only s-wave scattering occurs in the gas and one needs atoms belonging to two different hyperfine states in order to obtain a non-zero scattering, which is forbidden by Pauli principle between atoms belonging to the same hyperfine state. The scattering length $a$ varies accross the Feshbach resonance, starting with fairly small negative values at high field, which corresponds to a weakly attractive effective interaction between atoms belonging to two different hyperfine states. When the field is lowered $a$ goes to highly negative values and diverges right at the resonance, where it jumps an infinite positive value. When the field is further lowered the positive scattering length $a$ decreases down to fairly small values. On the low field side where $a>0$, molecules exist made of two fermionic atoms belonging to two different hyperfine states, while such a bound state does not exist on the $a<0$ side.

An experimental surprise has come from the fact that this singular behaviour, which occurs for two atoms in vacuum or for dilute Fermi gases, disappears when one goes to dense gases. For example the energy of the gas, measured in expansion experiments \cite{bourdel}, does not display any singularity and is perfectly smooth when one goes through the resonance, while the scattering length $a$ displays a singularity at the same location. This has been explained \cite{rc} by the effect of the dense Fermi gas, which introduces another length scale, namely the Fermi wavelength, related to the Fermi wavevector $k_F$, defined from the density $n$ of atoms belonging to a given hyperfine state (we assume that the density of the two hyperfine states are equal) $n=k_{F}^{3}/(6 \pi ^2)$. The associated energy scale is  the Fermi energy $E_F= k_F^2/(2m)$. The presence of the dense gas washes out the Feshbach resonance.

This effect is quite reasonable when one recalls that the very existence of the Cooper pairs, in the case of a very weak interaction, is due to the existence of the Fermi sea without which a bound state could not form. In other words the existence of the dense fermionic gas, or equivalently the Fermi sea, affects the formation of bound states, that is of molecules. In particular it has been shown explicitely that \cite{rcmol}, in the normal state, the threshold for the formation of molecules is affected by the presence of the dense gas. Instead of having, as in vacuum, all the molecular bound states appearing (with zero binding energy) at the same magnetic field corresponding to the Feshbach resonance, characterized by $a^{-1}=0$, the threshold for the appearance of these bound states depends now on the total momentum of the molecule. Qualitatively this effect is easily understood. Indeed, seen from a ${\bf k}$-space point of view, the formation of the molecular bound state requires the partial occupation of plane-wave states (the amplitude for the occupation probability is the Fourier transform of the molecular wave function). However in the presence of a dense gas, some of these states are already occupied and Pauli exclusion makes them unavailable for building up the molecular wave-function. As a result, with less plane-wave states available, the formation of the molecular bound state becomes more difficult and the threshold is naturally pushed toward stronger interactions, that is toward positive scattering lengths $a$.

It is also clear that this effect depends on the total momentum ${\bf K}$ of the molecule. If this momentum is very high, the molecular formation is unaffected since the plane-waves required to make up the wave function are near ${\bf K}$, and they are all essentially empty. In the case the threshold is at the same location as in vacuum, that is $a^{-1}=0$, independent of temperature. On the other hand for a molecule with ${\bf K}=0$ the required plane-waves have small wavevectors and these states are partially occupied. This is the case where the presence of the dense gas is the most strongly felt and the shift of the threshold is strongest, and this is the case we will more specifically consider. Naturally in all cases the effect depends on temperature since, when temperature is raised, the occupation of plane-wave states gets smaller and accordingly the effect on molecular formation is reduced. Ultimately at quite high temperature, the effect disappears because all the plane-wave states have very low occupation probability and we go back to the classical gas situation where the threshold is at $a^{-1}=0$ for any molecular state.

Naturally the experimental observation of this effect would be quite interesting, not the least because it goes against the simple intuition one gets from classical gas physics. This implies to have reasonable evaluations of the domain of physical parameters where it occurs. Unfortunately the evaluation made in Ref.\cite{rcmol} was very rough since the parameters of the Fermi sea were taken as those of a non interacting Fermi gas. This is naturally quite inconsistent since the appearance of molecules is due to interaction, which are in particular expected to have strong effects in the vicinity of the Feshbach resonance. This is even more so when one goes toward the BEC side, since one ends up with a system which is physically described as a dilute gas of molecular bosons, which has naturally little to do with a free Fermi sea. The initial purpose of the present paper is to proceed to a coherent calculation and describe the Fermi sea, taking into account self-consistently the interaction responsible for molecular formation.

More specifically the location of the molecular threshold has been obtained in Ref.\cite{rcmol} by writing that the full vertex for atomic scattering has a pole at zero energy $\omega =0$. This vertex has been obtained by a ladder approximation, which is a natural extension of the exact treatment for two atoms in vacuum. Here the existence of the dense gas has been taken into account by making use of the free fermion propagator for a given chemical potential $\mu $, instead of the fermionic propagator in vacuum. Now if we want to have consistently the properties of the fermionic gas, we should make use of this same vertex within the ladder approximation as a starting point. This is what is done in the present paper. From the vertex, we will obtain the fermionic self-energy, and then the fermionic density $n$ as a function of chemical potential $\mu $ and temperature $T$. In this way we will have in a consistent manner what is essentially the equation of state of the fermionic gas in the presence of interactions.

The appearance of the molecular bound state is not the only feature which is signaled by a pole in the vertex, the BCS transition appears also as a pole for an energy per particle equal to the chemical potential, so $\omega = 2 \mu $. This corresponds physically to the appearance of Cooper pairs, which are quite analogous to molecules. Hence it is a natural development to consider also the location of the BCS transition within our approximation. It is obviously of high interest to go toward the BEC side of the crossover and we will accordingly also determine the critical temperature for the Bose condensation of molecules, once we are beyond the threshold for their formation. All these features of the phase diagram, namely appearance of molecular bound states, BCS transition and Bose condensation of molecules, have the common property to be instabilities of the normal state and are revealed by the study of the same vertex.

As it happens the above approximation, we have been lead naturally to consider, has been already quite often considered in the literature in other contexts, dealing with interacting Fermi systems. It is often named the self-consistent $T$-matrix approximation and, as we mentionned above, it has been in particular used to handle the Hubbard model (see \cite{max1,max2} and references therein). In the context of ultracold fermionic gases, it has been already studied and used, in particular in the superfluid phases (see Ref.\cite{piestr,ppps,pps} and references therein). More specifically Pieri and Strinati \cite{piestr} obtain their equations as the result of a specific regularization for the well-known ultraviolet divergence which appears in the theory for a contact interaction. In their way the omission of diagrams other than ladder diagrams appears as an exact result. In our case we do not restrict ourselves to a contact interaction and keeping only ladder diagrams is just an approximation (physically quite reasonable), considered as to be possibly improved. Nevertheless we end up with the same equations, as can be more clearly seen by taking the normal state limit of the equations of Perali, Pieri, Pisani and Strinati \cite{ppps}.

On the other hand we will be concerned only by the normal state properties in the present paper, with more specifically in mind the study of the threshold for the molecular bound state, which might be relevant for quite recent experimental work \cite{vortex} as we will see. On the other hand their work \cite{ppps,pps} concentrates on the superfluid phase. Also they have focused on trapped gases, while we will deal only with the homogeneous gas. Actually the only overlap is at the level of the critical temperature of the superfluid phase, which they have also calculated for the homogeneous system \cite{ppps}. However our specific handling of the equations are different from theirs, each one having its own interest. More details and comparisons will be given in the course of our paper. Another specificity of our paper is that, in order to check the quality of our approximation, we will compare it in various non-trivial limiting cases with exact results already known in the literature. In all cases we will find a perfect agreement which establishes the self-consistent $T$-matrix approach as a physically very reasonable approximation. To our knowledge this specific comparison has not yet been done. Yet it is very important to assert precisely the quality of our description of the normal state. This is obviously quite necessary if we want to have a proper description of the superfluid state. This point has already been stressed quite a number of times in the context of high $T_c$ superconductivity, where it is often asserted that the normal state properties are even less properly understood than the superconducting ones. We stress that our handling of the equations will be made without any approximations, with either analytical or numerical treatment. In summary the present paper is complementary to the work of Pieri, Strinati and coworkers. In particular it shares the same spirit of trying to describe the physics of these ultracold gases starting from a single coherent fermionic picture for the whole crossover, in contrast to more phenomenologically oriented approaches where a bosonic field describing the molecular states is introduced immediately at the level of the Hamiltonian.

The organization of our paper is as follows. In the next section we introduce our basic equations to calculate the particle number in terms of the chemical potential and the temperature and we give explicitely the expressions for the self-energy and the vertex we will use throughout the paper. Then in the following section we consider the high temperature range and show that in this regime our approximation reduces to the exact virial expansion of Beth and Uhlenbeck \cite{beuh,LL} for the quantum gas. The interest of this limiting case has been emphasized recently by Pitaevskii and Stringari \cite{pitstr} who showed that it does not have any singular behaviour at the unitarity limit when the Feshbach resonance $a^{-1}=0$ is crossed, despite the appearance of molecular bound states. Then we show that, in the large momentum regime, at zero temperature, our approximation reduces to the known exact asymptotic result of Belyakov \cite{bely}. It is of particular experimental interest to emphasize this limiting case since very often the high momentum tail of the particle distribution is analyzed experimentally in order to extract the temperature, by comparison with the ideal case distribution, thereby ignoring the effects of interactions on this tail. We analyze also the dilute regime on the BCS side and we show how our approximation can be corrected to obtain perfect agreement with Galitskii's result \cite{gali}. Finally we show explicitely how the molecular bound state arises from the vertex and give the corresponding contribution to the self-energy. We analyze then the dilute limit on the BEC side of the crossover and show that one ends up, in the normal state, with the expected physical situation with the proper Bose distribution for molecules and the expected wave function for this nearly resonating situation. Finally we discuss in this regime another temperature, which is quite important in the context of high $T_c$ superconductivity, namely the pseudogap temperature $T_*$, which corresponds to the smooth crossover between the low temperature domain, where molecules (i.e. preformed pairs) dominate and the high temperature range where one has only free fermions. Then, gathering the different contributions to the self-energy, we calculate the momentum distribution for the atoms and total particle number. We emphasize in particular that a separation between particles corresponding to free atoms and particles belonging to a molecule has a quite restricted range of validity, although this analysis is quite often found in the literature in phenomenological analysis. Finally we end up by displaying and discussing our phase diagram for the molecular, the BCS and the Bose-Einstein instabilities.

\section{SELF ENERGY  AND  PARTICLE NUMBER}
\label{self}

The particle number $n$ in a single hyperfine state is obtained from the temperature Green's function $G({\bf k},i\omega _n)$, where $\omega _n = (2n+1) \pi T$ (with $n$ an integer) is the Matsubara frequency, by (we take $\hbar=1$ and $k_B=1$):
\begin{eqnarray}
n = T  \sum_{n}  \int \frac{d{\bf k}}{(2\pi )^{3}} \; G({\bf k},i\omega _n)\; e^{i \omega _n \tau}
\label{eqn}
\end{eqnarray}
where $\tau \rightarrow 0_+$. It is convenient, for actual calculations, to separate out in this equation the free particle contribution $G_0({\bf k},i\omega _n) = [i \omega _n - (\epsilon_{\bf k} - \mu)]^{-1}$, where $\epsilon_{\bf k} = k^{2}/2m$ is the free particle kinetic energy, for which the result $n_0$ for the particle number is known to be:
\begin{eqnarray}
n_0 = 4\pi  \int_{0}^{\infty} \frac{dk}{(2\pi )^{3}} \; \frac{k^2}{\exp (\frac{\xi_{\bf k} }{T}) + 1}
\label{eqn0}
\end{eqnarray}
where we have set $\xi_{\bf k} = \epsilon _k - \mu$. The remaining contribution is given by:
\begin{eqnarray}
n - n_0 = T  \sum_{n}  \int \frac{d{\bf k}}{(2\pi )^{3}} \; \frac{\Sigma (k, i\omega _{n})}{[i \omega _n - \xi_{\bf k}  - \Sigma (k, i\omega _{n})].[i \omega _n - \xi_{\bf k})]}
\label{eqnmn0}
\end{eqnarray}
since the Green's function is related to the self-energy $\Sigma (k, i\omega _{n}) $ by $G({\bf k},i\omega _n) = [i \omega _n - \xi_{\bf k}  - \Sigma (k, i\omega _{n})]^{-1}$.

In our ladder approximation the self-energy can be written as:
\begin{eqnarray}
\Sigma (k, i\omega _{n}) = T  \sum_{\nu}  \int \frac{d{\bf K}}{(2 \pi) ^{3}} \Gamma ({\bf K},i\omega _{\nu}) G_{0}({\bf K} - {\bf k},i\omega _{\nu}-i\omega _{n})
\label{sigmagen}
\end{eqnarray}
In contrast with the more general situation \cite{fetter} , we have here a single term because only atoms belonging to different hyperfine states do interact, which forbids an exchange term. Here $\Gamma ({\bf K},i\omega _{\nu})$ is the standard vertex \cite{fetter} in the ladder approximation. Note that in this vertex, $\omega _{\nu}= 2\pi \nu T$ (with $\nu$ being an integer) is a bosonic Matsubara frequency. Since the wavevector dependence of $\Gamma ({\bf K},i\omega _{\nu})$ is actually only on $K$, the angular integration of $G_{0}$ on ${\bf K}$ can be performed in Eq. (\ref{sigmagen}) and gives:
\begin{eqnarray}
 \int d\Omega _{\bf K} G_{0}({\bf K}- {\bf k},i\omega _{m}) =\frac{2 \pi m}{kK} \: \ln \frac{i\omega _{m} +\mu - K_{-}^{2}/2m}{i\omega _{m} +\mu - K_{+}^{2}/2m}
 \label{greenint}
\end{eqnarray}
with $K_{\pm} \equiv K \pm k$.

In the following we find it more convenient to work with reduced variables. We take as unit of energy the absolute value of the chemical potential $| \mu |$. This is the natural choice at low temperature in the degenerate regime, but not at high temperature in the classical regime where $T$ is a more convenient scale. Anyhow it is easy to switch in our formulae from one scale to another. Another inconvenience with our choice is that we have to keep track of the sign $s$ of $\mu $, defined by $s = \mu /| \mu | = \pm 1$, since it switches sign when one goes from the degenerate to the classical regime. We similarly define a wavevector scale $k_0$ by $k_{0}^{2}/2m = | \mu |$. Hence we set $\omega = | \mu | \,{\bar \omega }$ and so on for all the frequencies, together with $k = k_0 {\bar k}$ and $K = k_0 {\bar K}$. We also introduce the reduced temperature $t = T /| \mu |$.

The explicit expression for $\Gamma ({\bf K},i\omega _{\nu})$ is obtained, for example, as in Ref. \cite{rc}, but keeping a nonzero wavevector:
\begin{eqnarray}
\Gamma^{-1} ({\bf K},i\omega _{\nu}) = - \frac{mk_0}{2\pi ^2 \lambda} +  \int \frac{d{\bf k}'}{(2\pi )^{3}}
[ \,T  \sum_{m} G_0({\bf k}',i\omega _m) G_0({\bf K}-{\bf k}',i\omega _{\nu}-i\omega _m) -
 \frac{1}{2\epsilon _{k'}}\,]
\label{eqgam}
\end{eqnarray} 
where we have introduced the standard coupling constant $ \lambda = - 2 k_0 a/\pi $. The angular integration and the Matsubara summation can be performed and give, with the change of variable $k'=k_0x$:
\begin{eqnarray}
{\bar \Gamma}^{-1}({\bar K},{\bar \omega }) = \lambda^{-1} +  \int_{0}^{\infty} dx \left[1\,-\, \frac{xt/{\bar K}}{x^{2}+({\bar K}/2)^{2}-s-{\bar \omega }/2}\:\ln \frac{\cosh\left[((x+{\bar K}/2)^{2}-s)/2t\right]}{\cosh\left[((x-{\bar K}/2)^{2}-s)/2t\right]}\right]
\label{eqgamr}
\end{eqnarray}
where we have introduced a "reduced" vertex defined by ${\bar \Gamma}=-mk_0 \Gamma/(2\pi ^2)$
and used reduced variables.

In principle one could keep working with Matsubara frequencies and perform numerically the discrete frequency summation coming in the expression of the self-energy Eq.(\ref{sigmagen}). This is the procedure chosen for example by Perali \emph{et al} \cite{ppps}. However the corresponding series turn out to be rather slowly converging, which is numerically unpleasant. Hence we will rather transform here this summation into integrals over frequency in a standard way \cite{agd} by introducing the Fermi distribution $f(\omega /T) = 1/(\exp(\omega /T)+1)$ which has its poles at $ i \omega _n$, writing the summation as an integral over contours encircling these poles and deforming the contours. The contribution coming from the part of the contour at infinity is checked to be zero, and in general we are left with two contributions. We leave out for the moment a third possible contribution from a pole of $\Gamma$ on the real negative frequency axis, corresponding physically to a molecular state. This will be taken up in section \ref{mol}. Both of our contributions turn out to be rapidly convergent for large frequencies because of the Fermi distribution. One of them, which we call $ \Sigma_{\Gamma}$, arises from the cut of $\Gamma ({\bf K},i\omega _{n}+\omega) $ which extends from the branch point $ {\bar \omega _b} = 2 (({\bar K}/2)^{2}-s) - i {\bar \omega _{n}}$ to $\infty$. The other one $ \Sigma_{L}$ comes from the logarithm in Eq.(\ref{greenint}) and encircles the cut going from $ {\bar \omega _-}  = {\bar K}_{-}^{2}-s$ to 
$ {\bar \omega _+}  = {\bar K}_{+}^{2}-s$, where $ {\bar K}_{\pm} = {\bar K} \pm  {\bar k}$. For both contours we can rewrite their contributions as an integral along the corresponding cut by introducing the jump across the cut of the function to be integrated. We obtain in this way $\bar{\Sigma} (k, i\omega _{n}) \equiv  \Sigma (k, i\omega _{n}) / |\mu |= \bar{\Sigma}_{\Gamma} ({\bar k}, i{\bar \omega _{n}})  +\bar{ \Sigma}_{L} ({\bar k}, i{\bar \omega _{n}}) $ where:
\begin{eqnarray}
\bar{\Sigma}_{L}({\bar k},i{\bar \omega _{n}}) = - \frac{1}{2{\bar k}}  \int_{0}^{\infty} d{\bar K} \;{\bar K}\;  \int_{{\bar \omega _-} }^{{\bar \omega _+} } dy \; f(\frac{y}{t}) \; {\bar \Gamma}({\bar K},y+i {\bar \omega}_{n})
\label{eqsigl}
\end{eqnarray}
\begin{eqnarray}
\bar{\Sigma}_{\Gamma}({\bar k},i{\bar \omega _{n}}) =  - \frac{1}{2 \pi {\bar k}}  \int_{0}^{\infty} d{\bar K} \;{\bar K}\;  \int_{{\bar K}^{2}/2-2s}^{ \infty } dy \; b(\frac{y}{t}) \; \ln \left[ \frac{y-{\bar K}_{-}^{2}+s-i{\bar \omega}_{n}}{y-{\bar K}_{+}^{2}+s-i{\bar \omega}_{n}}\right]\; {\mathrm Im}{\bar \Gamma}({\bar K},y)
\label{eqsigg}
\end{eqnarray}
where ${\mathrm Im}{\bar \Gamma}({\bar K},y)$ is for ${\mathrm Im}{\bar \Gamma}({\bar K},y+i\epsilon )$ with $\epsilon \rightarrow 0_{+}$. We have introduced in the expression for $\bar{\Sigma}_{\Gamma}$ the Bose distribution $b(y) = 1/(\exp(y)-1)$.

\section{HIGH TEMPERATURE RANGE}

It is of interest to consider the high temperature limit of our approximation. We will see that it reduces to the virial expansion of Beth and Uhlenbeck \cite{beuh,LL} which means that it becomes exact in this limit. This is naturally quite a satisfactory feature of our approximation. For simplicity we restrict ourselves to the $a<0$ range where there is no molecular bound state. We have also proceeded to this comparison in the $a>0$ case, where the presence of bound states modifies the Beth and Uhlenbeck result, and we have found that our approximation reduces also to their result in this case.

In this limit we go to the classical regime where $ \mu \rightarrow -\infty$ with $  1/t = | \mu |/T \rightarrow \infty$ and in the above formulae we have $s=-1$. In this case the $\cosh$'s in Eq.(\ref{eqgamr}) may be replaced by exponentials and the integral on the right-hand side reduces to $(\pi /2) \sqrt{({\bar K}/2)^{2}+1-{\bar \omega}/2}$. This is just what one would get from the scattering amplitude of two particles in vacuum. This is naturally expected since, in this classical regime, one goes in a dilute limit. One can then see that the cut contribution $\bar{\Sigma}_{\Gamma}$ to the self-energy goes to zero as $ e^{-2| \mu |/T}$ because of the presence of the Bose distribution $b(y/t)$ and of the lower bound on the $y$ integration. It is therefore negligible compared to the pole contribution $\bar{\Sigma}_{L}$ which goes as $ e^{-| \mu |/T}$ as we will now see. Indeed in Eq.(\ref{eqsigl}) we may replace the Fermi distribution $f(y/t) = 1/(\exp(y/t)+1)$ by its classical limit $\exp(-y/t)$. Moreover, since $t \rightarrow 0$, $y$ is restricted in Eq. (\ref{eqsigl}) to a vanishingly small range above the lower bound ${\bar \omega _-}=({\bar K}-{\bar k})^2+1$, so in ${\bar \Gamma}$ we can replace $y$ by this lower bound. The $y$ integration is then easily performed. Similarly we want to have this lower bound as small as possible, in order to pick the dominant contribution from the Fermi distribution. This restricts ${\bar K}$ to a vanishingly small range around ${\bar k}$. Hence we can replace ${\bar K}$ by ${\bar k}$, except in the exponential, and the remaining integral is again easily performed. Finally it is more convenient to go back from our reduced variables to the physical ones, since in this limit the actual energy scale is $(2mT)^{1/2}$. This leads to:
\begin{eqnarray}
\Sigma (k, i\omega _{n}) = - \sqrt{\frac{2}{\pi }} \;T^{3/2} \; e^{-\frac{|\mu |}{T}} \frac{1}{-m^{-1/2}a^{-1}+ \sqrt{|\mu | -i\omega _n}}
= \frac{4\pi
n_0/m}{a^{-1}-\sqrt{m(|\mu|-i\omega_n)}}
\label{eqsight}
\end{eqnarray}
Actually we have omitted a term $(1/2) \epsilon _k$ in the square root, which turns out to be negligible since we will naturally find that only $\epsilon _k \sim T$ is relevant in our calculation, while $|\mu | \rightarrow \infty$. Hence the self-energy depends only on the frequency.

Let us first consider the case of small coupling $ |\lambda| \ll 1$, where the square root in the denominator is negligible compared to $m^{-1/2}a^{-1}$. In this case the self-energy is just a frequency-independent constant $\Sigma =a (2m/\pi)^{1/2} T^{3/2}  \exp (-|\mu |/T) $. Actually this result coincides, as it should, with the mean field expression $\Sigma = gn_0$ with the coupling constant $g=4 \pi a/m$. Hence this is just as if we had free particles, and had shifted the chemical potential from $\mu $ to $\mu - \Sigma$. The resulting change $ \delta n_k$ in the single hyperfine state particle distribution $n_k$ is, in this classical regime:
\begin{eqnarray}
 \delta n_k = - \frac{\Sigma}{T}\; e^{-\frac{\epsilon _k - \mu }{T}}
\label{eqnkwc}
\end{eqnarray}
After integration over ${\bf k}$, this leads to a change of particle density:
\begin{eqnarray}
\delta n = - \; \frac{\Sigma}{T} \;\Lambda_{T}^{-3}\; e^{-\frac{|\mu |}{T}}
\label{}
\end{eqnarray}
where we have introduced the thermal de Broglie wavelength $ \Lambda_T = (2 \pi /(mT))^{1/2}$. After substitution of the expression of $\Sigma$, this is $\delta n = - 2 a\,\Lambda_{T}^{-4} \,\exp (-2|\mu |/T)$, which is just the result of the virial expansion for our case.

Let us now come back to the general case where we have to take into account the frequency dependence of the self-energy. For simplicity we keep assuming that $a < 0$, so we have no bound state. Since the self-energy is small in this regime, the integrand in Eq. (\ref{eqnmn0}) reduces to $\Sigma (k, i\omega _{n})/(i \omega _n - \xi_{\bf k})^2$. The integration over $k$ is done first and gives $2i \pi (m/2)^{3/2}/ (i\omega _n - |\mu |)^{1/2}$ where the determination of the complex square root is with a positive imaginary part. On the other hand the square root $(|\mu |-i\omega _n)^{1/2}$ coming in Eq. (\ref{eqsight}) for the self-energy has the sign of its imaginary part opposite to the sign of $\omega _n$, as it can be seen from the starting expression Eq. (\ref{eqgamr}). Hence this square root has a cut on the real positive axis, starting from $\omega =|\mu |$.

The remaining summation over Matsubara frequencies is transformed in the standard way \cite{agd} into an integral over frequency on a contour encircling the poles of the Fermi distribution $f(\omega /T) = 1/(\exp(\omega /T)+1)$. This contour is then deformed into a contour $C$ which goes in a clockwise way around the cut $[|\mu |,\infty[$ of the square root \cite{note} , which gives:
\begin{eqnarray}
\delta n= - \frac{1}{2\pi ^2} \left(\frac{m}{2}\right)^{3/2}  \int_{C} d\omega \; \frac{\Sigma(\omega )}{e^{\omega /T}+1}
\; \frac{1}{(\omega - |\mu |)^{1/2}}
\label{eqhtinterm}
\end{eqnarray}
Taking into account that the determination of $(\omega - |\mu |)^{1/2}$ in the denominator changes sign when one crosses the cut $[|\mu |,\infty[$, this integral is twice the integral of its real part on the cut. Making the change of variable $\omega =|\mu |+p^2/m$ and using Eq. (\ref{eqsight}) for $\Sigma(\omega )$, we obtain finally in this classical limit $|\mu |/T \rightarrow\infty$:
\begin{eqnarray}
 \delta n = - \frac{1}{\pi} \: a \,\left(\frac{mT}{\pi }\right)^{3/2} e^{-2 |\mu |/T}\int_{0}^{\infty} dp \:
\frac{e^{-\frac{p^2}{mT}}}{1+p^2a^2}
\end{eqnarray}
which is just the Beth-Uhlenbeck result for our case. In the unitarity limit where $mTa^{2} \rightarrow \infty$ the integral reduces to $ \pi /(2 |a|)$ which gives $\delta n = (1/2) (mT/\pi )^{3/2}\,e^{-2 |\mu |/T}$. Naturally we have just written here the contribution coming from the interaction. The statistical correction \cite{beuh,LL} in the virial expansion will come out if we expand the Fermi distribution in Eq. (\ref{eqn0}).

\section{LARGE MOMENTUM BEHAVIOUR}
\label{gk}

Another limiting situation which is of interest to consider is the large momentum regime. A perturbative calculation in this limit has been done a long time ago by Belyakov \cite{bely}. Since his calculation was performed at $T=0$, we will also restrict ourselves to this case. Just as in the above section it is convenient to transform the summation over Matsubara frequencies in:
\begin{eqnarray}
n_k = T  \sum_{n} G({\bf k},i\omega _n)\; e^{i \omega _n \tau}
\label{eqnk}
\end{eqnarray}
into an integral over frequency on a contour encircling the poles of the Fermi distribution and to deform this contour into the contour $C=C_1 + C_2$ with $C_1=]i\infty-\epsilon,-i\infty-\epsilon[$ and $C_2= ]\!-i\infty+\epsilon,i\infty+\epsilon[$ with $\epsilon \rightarrow 0_+$. At $T=0$ the second part $C_2$ of this contour does not contribute, and we close the first part $C_1$ on the singularities of $G({\bf k},\omega)$ in the half-plane $\omega <0$.

For large ${\bar k}$, we see from Eq. (\ref{eqsigl}) that the bounds ${\bar \omega_{\pm}}=({\bar K}\pm {\bar k})^{2}-1$ are large, and the contribution to $\bar{\Sigma}_{L}$ negligible, except for the lower bound if ${\bar K}$ is very near ${\bar k}$. This makes ${\bar K}$ also large and, from Eq. (\ref{eqgamr}), one can use the limiting form ${\bar \Gamma}^{-1}({\bar K},y+{\bar \omega}) \approx \lambda^{-1}+(\pi /2) \sqrt{({\bar K}/2)^{2}-{\bar \omega}/2}$, since $y$ is bounded due to the Fermi distribution. The remaining integral is easily performed at $T=0$ to give:
\begin{eqnarray}
\bar{\Sigma}_{L}({\bar k},{\bar \omega}) = - \frac{2}{3} \; \frac{1}{\lambda^{-1}+(\pi /2) \sqrt{({\bar k}/2)^{2}-{\bar \omega}  }}
\label{eqsiglgk}
\end{eqnarray}
For ${\bar \omega}  <0$ this term in the self-energy has no singularity and does not generate any singularity in $G({\bf k},\omega)$. Hence we would merely have $n_k=0$ without the contribution of $\bar{\Sigma}_{\Gamma}$.

When $T \rightarrow 0$ the Bose distribution in Eq. (\ref{eqsigg}) restricts $y$ to negative value, which implies also from the lower bound of the integral that ${\bar K} \le 2$. Hence for large ${\bar k}$ the logarithm in this equation can be replaced by $-4{\bar k}{\bar K}/({\bar k}^2+i{\bar \omega}  _n)$. This gives for this case the following expression:
\begin{eqnarray}
\bar{\Sigma}_{\Gamma}({\bar k},{\bar \omega})= \frac{C}{{\bar \omega}+{\bar k}^2 }
\label{eqsigggk}
\end{eqnarray}
where the constant $C$ is given by:
\begin{eqnarray}
C=-\frac{2}{\pi } \int_{0}^{2} d{\bar K} \;{\bar K}^2\;  \int_{{\bar K}^{2}/2-2}^{0} dy \; \; {\mathrm Im}{\bar \Gamma}({\bar K},y)
\label{eqc}
\end{eqnarray}
This constant is easily evaluated since the imaginary part of ${\bar \Gamma}^{-1}({\bar K},y)$ becomes simple from Eq. (\ref{eqgamr}) at $T=0$ and, if we restrict ourselves to the perturbative situation of small $\lambda$ investigated by Belyakov, its real part is merely $\lambda^{-1}$. This leads to $C=4\lambda^2/9$.

Since $\bar{\Sigma}_{\Gamma}$ has a pole for ${\bar \omega} =-{\bar k}^2$, it is easy to see that $G({\bf k},\omega)$ itself has a pole in the vicinity. In this vicinity $\bar{\Sigma}_{L}$ is small and can be neglected. This leaves us with:
\begin{eqnarray}
G^{-1}({\bf k},\omega)=|\mu | [ {\bar \omega} - {\bar k}^2 - \frac{C}{{\bar k}^2+{\bar \omega}}] \simeq \frac{-2|\mu |{\bar k}^2}{{\bar k}^2+{\bar \omega}}\; [{\bar \omega}+{\bar k}^2 +\frac{C}{2{\bar k}^2}]
\label{eqgkgm}
\end{eqnarray}
The pole of $G({\bf k},\omega)$ at ${\bar \omega}=-[{\bar k}^2 +C/2{\bar k}^2]$ has a residue $C/(4|\mu |{\bar k}^4)$ which leads to:
\begin{eqnarray}
n_k=\frac{C}{4 {\bar k}^4}=\left(\frac{2}{3\pi }\,k_0 a\right)^{2} \;\frac{k_0^4}{k^4}\label{eqnk}
\end{eqnarray}
in agreement with Belyakov \cite{bely}. One would also obtain this result by performing straight away an expansion in the small parameter $C$, but this is not so obvious to justify since the denominator ${\bar k}^2+{\bar \omega}$ in Eq. (\ref{eqgkgm}) can vanish.

\section{DILUTE LIMIT FOR NEGATIVE SCATTERING LENGTH}
\label{dilbcs}

The result of Belyakov has been obtained by second order perturbation theory, and the agreement we find with his result would let us believe that our approximation is completely valid up to second order in perturbation. This is actually not completely correct, as it can be seen by comparing our results with Galitskii's dilute limit theory \cite{gali,fetter}. We just sketch here the calculation, at $T=0$, which is basically an expansion in powers of the scattering length $a$ of the expression of the particle density $n$ in terms of the chemical potential $\mu $.

If we insert in Eq. (\ref{eqn}) the first order expression $\Sigma^{(1)}=gn_0$ where $g=4\pi\,a/m$, we have for the particle density:
\begin{eqnarray}
n^{(1)}&=& \int \frac{d{\bf k}}{(2\pi)^3} \int \frac{d\omega}{2\pi} \frac{e^{i\omega 0_{+}}}{(i\omega-\xi_{\bf k} -g\,n_0)}
\label{eqngal1}
\end{eqnarray}
As is well known, this is just the zeroth order result $n_{0}(\mu')$, with a shifted chemical potential 
$\mu'=\mu-g\,n_0$. To proceed, we consider the difference $n-n^{(1)}$  
and we evaluate the self-energy up to second order $\Sigma(k,i\omega) - g\,n_0\simeq\Sigma^{(2)}(k,i\omega)$. Similarly to Eq. (\ref{eqnmn0}) this gives:
\begin{eqnarray}
\label{eqngal2}
n-n^{(1)}&=&\int \frac{d{\bf k}}{(2\pi)^3}\int\frac{d\omega}{2\pi}\frac{\Sigma^{(2)}(k,i\omega)}
{(i\omega-\xi'_{\bf k})(i\omega-\xi'_{\bf k} -\Sigma^{(2)}(k,i\omega))}
\end{eqnarray}
where we have set $\xi'_{\bf k} = \epsilon _k - \mu'$.
If one sets directly $\Sigma^{(2)}=0$ in the denominator, one finds $n-n^{(1)}=0$. In order to get the proper expansion, we substract from Eq. (\ref{eqngal2}) the corresponding expression
without $\Sigma^{(2)}$ in the denominator, which leads to: 
\begin{eqnarray}
\label{eqngal3}
n-n^{(1)}&=&\int \frac{d{\bf k}}{(2\pi)^3}\int\frac{d\omega}{2\pi}\;
\frac{\Sigma^{(2)}(k,i\omega)}{i\omega -\xi'_{\bf k}}
\left[\frac{1}{i\omega -\xi'_{\bf k} -\Sigma^{(2)}(k,i\omega)}-\frac{1}{i\omega-\xi'_{\bf k}}
\right]
\end{eqnarray}
We see that the integrand is important only when $i\omega -\xi'_{\bf k}$ is small,
{\it i.e.} when the variables are close to $\omega=0$ and $k=k'_0$, where $\xi'_{\bf k'_0}=0$. Therefore, one may replace in the integral $\Sigma^{(2)}(k,i\omega)$ by
 $\Sigma^{(2)}(k'_0,0)=C'\,(k'_{0}a)^2\,\mu$, where $C'$ is a positive constant to be evaluated, and consider a small domain of integration around
$\omega=0$ and $k=k'_0$. Performing the integrations gives $n-n^{(1)}=-\frac{k_{0}^3}{4\pi^2}(k_0 a)^2 \,C'$, where we have made in this second order term $k'_0 \approx k_0$. We have therefore, for $n\equiv \frac{k_{F}^3}{6\pi^2}$, the second order expansion we were looking for:
\begin{eqnarray}  
\frac{k_{F}^3}{6\pi^2}&=&\frac{1}{6\pi^2}\left(2m(\mu-g\,n_0)\right)^{3/2}
-\frac{C'}{4\pi^2}k_0^3\,(k_0\, a)^2+\cdots 
\label{eqkf}
\end{eqnarray}
where $k_0=\sqrt{2m\mu }$ and $n_0=k_0^3/(6\pi ^2)$.

The constant $C'$ is determined by considering the second order values of the self-energies contributions
$\Sigma_L^{(2)}$ and $\Sigma_\Gamma^{(2)}$. 
Performing analytically the integrals for $\Sigma_\Gamma^{(2)}$, one finds
$\Sigma_\Gamma^{(2)}(k_0,0)=(2+\log 2) \,\frac{8}{15\pi^2} \,\frac{k_0^2}{2m}(k_0 a)^2 $. 
The integrals entering $\Sigma_L^{(2)}$ are more easily calculated numerically and we get 
$\Sigma_L^{(2)}(k_0,0)=0.2818 \, \frac{4}{\pi^2} \frac{k_0^2}{2m}(k_0 a)^2 $. 
From this we obtain the numerical value $C'=0.2597$. Inverting the expansion in Eq. (\ref{eqkf}) in order to express the
chemical potential in powers of $k_{F}\,a$, we have:
\begin{eqnarray}
\label{eqgal}
\mu=\frac{k_{F}^2}{2m}\left(1+\alpha\,k_{F}\,a+\beta\,(k_{F}\,a)^2+\cdots\right)
\end{eqnarray}
We get $\alpha = \frac{4}{3\pi}$ and $\beta=\frac{8}{3\pi^2}+C'$.
This is in apparent disagreement with Galitskii's result where $\beta=\frac{4}{15\,\pi^2}(11-2\log 2)\simeq 0.2597$.

This is due to the fact that our self-consistent calculation does not include all the second order contributions. Indeed when we consider the Hartree term $gn$, we see that we have taken it into account only to zeroth order, in $\Sigma^{(1)}=gn_0$. However if we want to have a result which is valid up to second order, we want to have the relation $n(\mu)$ in $gn$ correct up to first order while we have it only to zeroth order. Technically this is because we do not have any self-energy contribution in the propagator $G$ which comes in the familiar diagram which corresponds to the Hartree term, that is we take $G=G_0$ in this diagram, instead of taking into account a first order correction to $G$. We note that this defect of our scheme is easily corrected "by hand": since we know that the exact result for the Hartree term (to any order) is anyway  $gn$, proportional to the actual density, we can make directly this correction in our formalism. Indeed if we replace in Eq. (\ref{eqkf}) $gn_0$ by $gn$, we obtain $\beta=C'$, in full agreement with Galitskii.

\section{MOLECULAR BOUND STATE}
\label{mol}

As mentionned in section III, we have not yet considered the possibility that $\Gamma ({\bf K},\omega)$ has a pole on the real frequency axis, corresponding physically to the existence of a bound state formed by two fermions belonging to two different hyperfine states, in other words a molecular state. Naturally this will occur only for $a>0$. This is what we will study in the present section.

This pole will appear below the branch point for the cut, which is at ${\bar K}^2/2-2s$. Let us call $y_0({\bar K})$ the corresponding zero occuring in Eq. (\ref{eqgamr}). The equation for $Y_0({\bar K})=y_0({\bar K})-{\bar K}^2/2+2s$, which is negative, is given explicitely by:
\begin{eqnarray}
\lambda^{-1} +  \int_{0}^{\infty} dx \left[1\,-\, \frac{xt/{\bar K}}{x^{2}-Y_0/2}\:\ln \frac{\cosh\left[((x+{\bar K}/2)^{2}-s)/2t\right]}{\cosh\left[((x-{\bar K}/2)^{2}-s)/2t\right]}\right] =0
\label{eqY0}
\end{eqnarray}For increasing coupling strength this zero will first appear when it is right at the branch point, that is $Y_0=0$. This gives for the corresponding coupling strength threshold $\lambda_{th}({\bar K})$: 
\begin{eqnarray}
- \lambda_{th}^{-1}({\bar K}) = \int_{0}^{\infty} dx \left[1\,-\, \frac{t}{x{\bar K}}\:\ln \frac{\cosh\left[((x+{\bar K}/2)^{2}-s)/2t\right]}{\cosh\left[((x-{\bar K}/2)^{2}-s)/2t\right]}\right]
\label{eqlambseuil}
\end{eqnarray}
As expected this coupling strength is always negative, since one can check that the quantity to be integrated in Eq. (\ref{eqlambseuil}) is positive. This result gives the threshold for formation of a molecule with total momentum $K = k_0 {\bar K}$. In particular for zero momentum ${\bar K}=0$, one obtains:
\begin{eqnarray}
- \lambda_{th}(0)^{-1} = \int_{0}^{\infty} dx \left[1\,-\, \tanh \frac{x^2+1}{2t}\right]
\label{eqlambseuilv0}
\end{eqnarray}
which has already been studied in Ref. \cite{rcmol}. In this case we have taken $s=-1$ since otherwise one is already in the phase diagram domain where the BCS instability occurs \cite{rcmol}. On the other hand for ${\bar K} \neq 0$ both $s= \pm 1$ are possible. Finally one checks that, for large ${\bar K}$ Eq. (\ref{eqY0}) gives merely $y_0 = {\bar K}^2/2-2s - 8/(\pi \lambda)^2$. In  particular the threshold is given by $\lambda_{th}^{-1}({\bar K}) = 0$. This is expected since in this case atoms become insensitive to the presence of the other fermions, and one should recover the result for atoms in vacuum.

The existence of a bound state implies that we have to consider in Eq. (\ref{sigmagen}) the additional contribution $\bar{\Sigma}_{m}$ coming from the corresponding pole occuring at $y_0$. It is given by:
\begin{eqnarray}
\bar{\Sigma}_{m}({\bar k},i{\bar \omega _{n}})=-\frac{1}{2{\bar k}} \int_{0}^{\infty} d{\bar K} \;{\bar K}\; b(\frac{y_0}{t}) \;  
\ln \left[ \frac{y_0-i{\bar \omega } _n-{\bar K}_{-}^{2}+s}{y_0-i{\bar \omega } _n-{\bar K}_{+}^{2}+s}\right]\;
\left[\frac{\partial {\bar \Gamma}^{-1}}{\partial {\bar \omega} }\right]_{y_ 0}^{-1}
\label{eqsigm}
\end{eqnarray}
Introducing:
\begin{eqnarray}
J({\bar K})= \int_{0}^{\infty} dx \, \frac{xt/{\bar K}}{[x^{2}+|Y_0/2|]^2}\:\ln \frac{\cosh\left[((x+{\bar K}/2)^{2}-s)/2t\right]}{\cosh\left[((x-{\bar K}/2)^{2}-s)/2t\right]}
\label{eqgj}
\end{eqnarray}
we can rewrite the above contribution as:
\begin{eqnarray}
\bar{\Sigma}_{m}({\bar k},i{\bar \omega _{n}})=-\frac{1}{{\bar k}} \int_{0}^{\infty} d{\bar K} \;{\bar K}\; b(\frac{y_0}{t}) \;  
\frac{1}{J({\bar K})}\;\ln \left[ \frac{y_0-{\bar K}_{-}^{2}+s-i{\bar \omega _{n}}}{y_0-{\bar K}_{+}^{2}+s-i{\bar \omega _{n}}}\right]\;
\label{eqsigm1}
\end{eqnarray}

\section{DILUTE LIMIT FOR POSITIVE SCATTERING LENGTH}
\label{dilbec}

We explore now the behaviour of our approximation on the BEC side of the phase diagram, going to the dilute limit. This corresponds to take $a>0$ small enough. In this case we will have the presence of molecular bound states, corresponding to the pole of the vertex considered in the preceding section. The binding energy becomes large when $a$ is small. Since physically the chemical potential will be at most half the bound state energy, it will be large and negative in this range. This implies that $t=T/|\mu | \ll 1$ and $s=-1$. More specifically in this case, the arguments of the cosh's in Eq. (\ref{eqY0}) are always large and we find $|Y_0|=2 (2/\pi \lambda)^{2}$. For a molecule at rest (i.e. ${\bar K}=0$) this means that the molecular binding energy is $|\mu |.|Y_0|=\epsilon _b=1/(ma^2)$ as expected. Similarly, from Eq. (\ref{eqgj}), the expression of $J({\bar K})$ simplifies into $J({\bar K})=\pi /(4 \sqrt{|Y_0/2})=\pi ^2|\lambda|/8$.

Turning now to the self-energy we see from Eq. (\ref{eqsigl}) and Eq. (\ref{eqsigg}) that, because of the Fermi and Bose distributions, these contributions $\bar{\Sigma}_{L}$ and $\bar{\Sigma}_{\Gamma}$ to the self-energy will contain exponentially small factors, respectively $\exp(-1/t)$ and $\exp(-2/t)$, since the lower bound for the $y$ integral is larger than (or equal to) 1 or 2, respectively. Hence we can neglect these two contributions in this small $t$ limit, and retain only the contribution $\bar{\Sigma}_{m}$ considered in the preceding section, which can be checked at the end of our calculation to be indeed exponentially larger than the two other ones.

The expression Eq. (\ref{eqsigm1}) for $\bar{\Sigma}_{m}$ simplifies since the Bose distribution factor forces ${\bar K}$ to be very small, and the argument of the logarithm is very near unity. This leads to:
\begin{eqnarray}
\bar{\Sigma}_{m}({\bar k},{\bar \omega})=\frac{C}{{\bar \omega}+|Y_0|+{\bar k}^{2}-1}
\label{eqsigm2}
\end{eqnarray}
where the constant:
\begin{eqnarray}
C=\frac{16}{\pi}\sqrt{\frac{|Y_0|}{2}}\; \int_{0}^{\infty} d{\bar K} \;{\bar K}^2\; b(\frac{y_0}{t})
\label{eqgc}
\end{eqnarray}
is small in the dilute regime we are interested in.

The expression for the self-energy is made clearer by going back to physical variables and making use of Eq.(\ref{eqnbose}) below,which gives:
\begin{eqnarray}
\Sigma_m(k,i\omega_n)=\frac{16\pi
a|\mu|n/m(k_0a)^2}{i\omega_n+\xi_k+\mu_B}
\label{eqsigm3}
\end{eqnarray}
where we have set $\mu_B=2\mu+\epsilon _b$, which we will justify physically just below. Note that $\Sigma_m(k,i\omega_n)$ has a "hole"-like dispersion in the denominator
$i\omega_n+\xi_k+\mu_B$ in contrast to the "particle"-like
dispersion $i\omega_n-\xi_k$ in the bare Green function $G_0$.

The situation is now similar to the one we had in section \ref{gk} for the large momentum behaviour. Since we have $t \ll 1$ (which does not imply that $T$ itself goes to zero), once again only the singularities of $G({\bf k},\omega)$ in the half-plane $\omega <0$ will contribute in the calculation of $n_k$ given by Eq. (\ref{eqnk}). The self-energy Eq. (\ref{eqsigm2}) has a pole for ${\bar \omega}=1-|Y_0|-{\bar k}^{2}$ and we find that $G({\bf k},\omega)$ itself has a pole in its vicinity at ${\bar \omega}= 1-|Y_0|-{\bar k}^{2} -C/(|Y_0|+2{\bar k}^{2})$. The corresponding residue is $C/(|Y_0|+2{\bar k}^{2})^2$. This leads to:
\begin{eqnarray}
n_k= \frac{C}{(|Y_0|+2{\bar k}^{2})^2}
\label{eqnkbose}
\end{eqnarray}
and after integration over ${\bar k}$ we have finally for the total density for a single hyperfine state:
\begin{eqnarray}
\frac{n}{k_0^3}=\frac{C}{2\pi ^2}  \int_{0}^{\infty} d{\bar k} \;\frac{{\bar k}^2}{(|Y_0|+2{\bar k}^{2})^2}=
\frac{C}{32 \pi} \sqrt{\frac{2}{|Y_0|}}=\frac{C}{32\pi}k_0a
\label{eqnbose}
\end{eqnarray}

These results Eq. (\ref{eqnkbose}) and Eq. (\ref{eqnbose}) are just what one expects in this regime. Indeed taking Eq. (\ref{eqgc}) into account and going back from reduced to physical variables we can rewrite Eq. (\ref{eqnbose}) as:
\begin{eqnarray}
n= \; \frac{4\pi }{(2\pi )^3}  \int_{0}^{\infty} dk \;\frac{k^2}{\exp[(\frac{k^2}{4m}-\epsilon_b -2\mu ) /T]-1} 
\label{eqnbose1}
\end{eqnarray}
which is just what is expected for the density of non interacting bosonic molecules with binding energy $\epsilon _b$, mass $2m$ and chemical potential $\mu _B=2\mu +\epsilon _b $. Similarly we know that the wavefunction for the relative motion of atoms in these large nearly resonating molecules is $\psi (r)=(A/r) e^{-r/a}$, where $A$ is a normalization constant. The momentum distribution is proportional to the square of the Fourier transform of this wavefunction, that is $(k^2+1/a^2)^{-2}$, which is just what we find in Eq. (\ref{eqnkbose}) in reduced variables, since $|Y_0|=\epsilon _b/|\mu |=2/(k_0a)^2$. It is worthwhile to note that we are in the normal state, in contrast with the similar result obtained from the approximate BCS wave function in the strong coupling limit, which holds in the superfluid state.

We may wonder if our calculation could also produce interactions between our molecular states. Although this is not so obvious, this does not seem to be the case. Indeed in our expression Eq. (\ref{eqsigm1}) for the self-energy, we have only a single Bose distribution which is appearing, which merely leads to the expression Eq. (\ref{eqnbose}) for the density where this Bose factor appears again. Hence if we had interactions we would expect the appearance of products of Bose distributions, which we do not have. In a related way we notice that this Bose factor is linked to the existence of ladders in the diagrammatic writing of our formulation. This corresponds physically to the propagation of a single molecule. However in our formulation there are no diagrams where two ladders interact. Physically this means that we have no interactions between molecules. In order to find them one has to consider more complicated diagrams, describing such an interaction, as it has been done for example by Pieri and Strinati \cite{piestr} and very recently by Brodsky \emph{et al} \cite{bkkcl}, giving an exact agreement with the four fermions calculation of Petrov \emph{et al} \cite{petrov}.

\section{TEMPERATURE EVOLUTION OF THE SYSTEM IN THE MOLECULAR LIMIT} 
\label{sec_eight}
In this dilute molecular limit
$\epsilon _b=1/(ma^2) \gg E_F$ it is
interesting to consider in more details the temperature evolution of
the system. At high temperatures $T \gg E_F$ the situation
is governed here by the dynamical equilibrium between molecules
and unbound fermions. This situation is described by the
well-known Saha (or law of mass action) formula (see \cite{LL}). In 3D it reads:
\begin{eqnarray}
\frac{n^2_{F}}{n_B}=\frac{1}{2\pi^2}(mT)^{3/2}\exp(-\epsilon _b/T),
\label{eqn_k1}
\end{eqnarray}
where the total particle density ${\tilde n}=2n=n_F+2n_B$, with $n_F$ the sum of the free fermion density  and $n_B$ the bosonic molecular density. The crossover temperature $T_*$ for which $n_F=2n_b={\tilde n}/2$
is given by:
\begin{eqnarray}
T_*\approx\frac{E_b}{\frac{3}{2}\ln\frac{\epsilon _b}{E_F}},
\label{eqn_k2}
\end{eqnarray} where the logarithm in denominator of Eq.(\ref{eqn_k2}) has an entropic
character. For $T \gg T_*$ one has $ n_F \gg 2n_B$, hence ${\tilde n}\approx n_F $ and the
fermionic chemical potential
$\mu\approx-\frac{3}{2}T\ln(T/T^{BEC}_c)$ has the standard
Boltzman form ($T^{BEC}_c \sim E_F$ is a typical temperature for Bose-Einstein
condensation).

At much lower temperatures $E_F \ll T \ll T_*$ the situation
drastically changes. The number of unpaired fermions
$n_F\sim\exp(-\epsilon _b/2T)$ becomes exponentially small and hence
${\tilde n}\approx2n_B$. The fermionic chemical potential acquires a kink:
\begin{eqnarray}
\mu\approx-\frac{\epsilon _b}{2}-\frac{3}{4}T\ln\frac{T}{T^{BEC}_c}
\label{eqn_k3}
\end{eqnarray}
In this low temperature regime we have essentially $2\mu = -\epsilon _b$, that is $k_0a=1$. This simplifies the expression Eq.(\ref{eqsigm3}) into:
\begin{eqnarray}
\Sigma_m(k,i\omega_n)=\frac{16\pi
a|\mu|n/m}{i\omega_n+\xi_k+\mu_B}
\label{eqsigm4}
\end{eqnarray}

It is important to emphasize that for $T \ll T_*$ our two-particle
vertex has the simple pole structure:
\begin{eqnarray}
\Gamma({\bf K},i\omega_\nu)=\frac{4|\mu|\frac{4\pi
a}{m}}{i\omega_\nu-\frac{K^2}{4m}+\mu_B}, \label{eqn_k4}
\end{eqnarray} where, as we have seen in the preceding section, $\mu_B=2\mu+\epsilon _b$ is the molecular chemical potential. Correspondingly we have $\mu_B \approx-\frac{3}{2}T\ln\frac{T}{T^{BEC}_c}$.

The dressed one-particle Green function
$G^{-1}({\bf k},i\omega_n)=G^{-1}_0({\bf k},i\omega_n)-\Sigma({\bf k},i\omega_n)$
has a two-pole structure for $T \ll E_F$:
\begin{eqnarray}
G=\frac{1}{i\omega_n-\xi_k-\frac{16\pi
a|\mu|n/m}{i\omega_n+\xi_k+\mu_B}}, \label{eqn_k5}
\end{eqnarray}where, as before, $\xi_k=k^2/2m-\mu$.

The spectral function $A({\bf k},\omega)=-\frac{1}{\pi}\,{\mathrm Im}G({\bf k},\omega+i0_{+})$ reads:
\begin{eqnarray}
A({\bf k},\omega)\approx(1-\frac{4\pi
a|\mu|n/m}{\xi^2_k})\delta(\omega-\xi_k)+\frac{4\pi
a|\mu|n/m}{\xi^2_k}\delta(\omega+\xi_k+\mu_B)
\end{eqnarray}
It reflects for $T \ll T_*$ the existence of two bands: the filled
bosonic band and, separated by the correlation gap $\epsilon _b$, the
almost empty band of unbound fermions. Integrating the spectral weight it is easy to check,
that in this regime: 
\begin{eqnarray}
\frac{4\pi a|\mu|n}{m}\frac{1}{2\pi^2}\int_0^{\infty}\frac{k^2dk}{\xi_k^2}=n \approx 2n_B
\label{eqweight}
\end{eqnarray}

The specific heat of the system:
\begin{eqnarray}
C_v=\frac{\partial E}{\partial T}=\frac{\partial }{\partial
T}[\int\frac{k^2}{4m}\frac{k^2dk}{2\pi^2}\exp(-\frac{k^2}{4mT})\exp\frac{\mu_B}{T}]\sim
n={\mathrm const.} \label{eqn_k6}
\end{eqnarray} is temperature independent in agreement
with general thermodynamic requirements.

\section{GENERAL MOMENTUM DISTRIBUTION}
\label{mom}

An intermediate step in our calculation is naturally the particle momentum distribution $n_k=n_{0k}+\delta n_k$, where $n_{0k}=(\exp(\xi_k/T)+1)^{-1}$ is the Fermi distribution and the correction due to interactions is, with our reduced units, given by:
\begin{eqnarray}
\delta n_k = 2\,t\; {\mathrm Re} \sum_{n=0}^{\infty} \; \frac{\bar{\Sigma }({\bar k}, i\bar{\omega} _{n})}{[i \bar{\omega} _n - {\bar k}^2 +s  - \bar{\Sigma} ({\bar k}, i\bar{\omega} _{n})].[i \bar{\omega} _n - {\bar k}^2 +s]}
\label{eqdnk}
\end{eqnarray}
with $k=k_0 {\bar k}$ and we have used $\Sigma (k, -i\omega _{n})=\Sigma^{*} (k, i\omega _{n})$, and we have explicitely in terms of the contributions considered above $\bar{\Sigma }({\bar k}, i\bar{\omega} _{n})=\bar{\Sigma }_{\Gamma}({\bar k}, i\bar{\omega} _{n})+\bar{\Sigma }_{L}({\bar k}, i\bar{\omega} _{n})+\bar{\Sigma }_{m}({\bar k}, i\bar{\omega} _{n})$. Then the particle number is obtained by $n=(1/2\pi^2)  \int_{0}^{\infty} \!dk\,k^2\,n_k$. The calculations of $\delta n_k$ and of $n$ are both handled numerically.

The above Matsubara summation is numerically quite convenient since it converges fairly rapidly (for large $\bar{\omega} _n$ the terms in the series Eq. (\ref{eqdnk}) behave typically as $1/\bar{\omega} _n^{3}$). Nevertheless it is of interest to consider another possible calculation where this summation would be transformed in a contour integration over frequency as we have done for example in section \ref{gk} and \ref{dilbec}. The contour would be deformed to enclose the singularities of the Green's function occuring on the real frequency axis, and the result could be expressed in terms of the spectral density. This would allow to ascribe a physical meaning to the various contributions, as we have done in section \ref{gk} where we had to deal with a fermion pole, whereas in section \ref{dilbec} we had a molecular pole. This is of interest since the total particle number $n$ is very often split in the literature into a free fermion term plus a molecular term as $n=n_{\mathrm ferm}+n_{\mathrm mol}$. 

It is easy to see that this split is not possible in general because the Green's function has a cut on the real axis extending from $-\infty$ to $+\infty$. There is no way to split the Green's function itself in a sum of a free fermion term and a molecular term since this is rather at the level of the self-energy that such a separation occurs. We have indeed found for the self-energy a molecular contribution $\Sigma_m$, linked to the molecular pole of $\Gamma ({\bf k},\omega)$. The contribution $\Sigma_{\Gamma}$ can also be understood as linked to molecules (as it is clear from the Bose factor it contains in Eq. (\ref{eqsigg})), but this is rather the continuous spectrum of molecules broken into two fermions, rather than the discrete spectrum linked to $\Sigma_m$. Finally the contribution $\Sigma_L$ is linked to single fermions since it arises in Eq.(\ref{sigmagen}) from the pole of the Green's function $G_0$ which is clearly signaled by the Fermi factor in Eq.(\ref{eqsigl}). We could still think of separating a molecular contribution from a fermionic contribution if we had in the spectral density of the Green's function an energetically well separated part, similar to what we had in section \ref{dilbec}. However this is likely to occur only when the molecular binding energy is large compared to temperature, in which case we will have dominantly molecules, and the separation of the density into a molecular part and a fermionic part is rather uninteresting. Anyhow it is clear from the above discussion that such a separation is in general unwarranted and, if it is used, it must taken cautiously.

\section{COMPARISON  WITH  OTHER  WORKS}
\label{comp}

As we have already indicated our approach is in complete agreement with the equations written by Perali, Pieri, Pisani and Strinati \cite{ppps} when we restrict them to the normal state. It is now also of interest to compare our framework with the one used by Nozi\`eres and Schmitt-Rink \cite{nsr} (NSR). 

Quite generally we can write \cite{agd} the thermodynamic potential (function of $\mu $ and $T$) of the two hyperfine states to as:
\begin{eqnarray}
\Omega - \Omega_0 =  \int_{0}^{\lambda}  \frac{d\lambda '}{\lambda'} \;T  \sum_{{\bf k},\omega_n}
G_{0}^{-1}({\bf k},\omega_n) [ G({\bf k},\omega_n)- G_0({\bf k},\omega_n) ]
\label{eqgpot}
\end{eqnarray}
where the integration is over a varying coupling constant $\lambda'$. Here $\Omega_0$ is the thermodynamic potential for the non interacting gas, $G({\bf k},\omega_n)$ and $G_0({\bf k},\omega_n)$ are the Green's functions for the interacting and the non interacting gas. Introducing the expansion of the full Green's function in terms of the self-energy $ G = G_0  \sum_{p=0}^{\infty} (\Sigma G_0)^p$, this can be rewritten as:
\begin{eqnarray}
\Omega - \Omega_0 =  \int_{0}^{\lambda}  \frac{d\lambda '}{\lambda'} \;T  \sum_{{\bf k},\omega_n} \sum_{p=1}^{\infty} \left[\Sigma({\bf k},\omega_n) G_0({\bf k},\omega_n)\right]^p
\label{eqgpot1}
\end{eqnarray}
In our approximate scheme the self-energy is given by Eq.(\ref{sigmagen}), with $ \Gamma =-2\pi ^2 {\bar \Gamma}/(mk_0)$ and ${\bar \Gamma}^{-1}=\lambda^{-1}+I$ from Eq. (\ref{eqgamr}), where $I$ denotes the integral in the right-hand side of this equation. If we consider the first order term $ \Sigma G_0$ in the above expansion we can rewrite its contribution as:
\begin{eqnarray}
T  \sum_{{\bf k},\omega_n} \Sigma({\bf k},\omega_n) G_0({\bf k},\omega_n) = T  \sum_{{\bf k'},\omega'_n} \frac{\lambda I({\bf k'},\omega'_n)}{1+\lambda I({\bf k'},\omega'_n)}= T  \sum_{{\bf k'},\omega'_n}  \sum_{p=1}^{\infty}(-1)^{p-1}[\lambda I({\bf k'},\omega'_n)]^p
\label{eq1st}
\end{eqnarray}
where $\omega'_n$ is a bosonic Matsubara frequency. To obtain this equation we have used $ T  \sum_{{\bf k},\omega_n} G_0({\bf k},\omega_n)G_0({\bf k}'-{\bf k},\omega'_n-\omega_n) = -mk_0  I({\bf k'},\omega'_n)/(2\pi ^2)$, as it results from comparing Eq.(\ref{eqgam}) and Eq.(\ref{eqgamr}), and we have omitted the regularization term of the high momentum divergence, since NSR use rather a separable potential to achieve this regularization. Inserting this contribution in Eq.(\ref{eqgpot1}) we find
for the corresponding contribution to the thermodynamic potential:
\begin{eqnarray}
\label{eqgpotnsr}
\Omega - \Omega_0 = T  \sum_{{\bf k},\omega_n} \sum_{p=1}^{\infty}\frac{(-1)^{p-1}}{p}[\lambda I({\bf k},\omega_n)]^p = T  \sum_{{\bf k},\omega_n} \ln [1+\lambda I({\bf k},\omega_n)]
\end{eqnarray}
which is just NSR expression, provided that we identify their $\chi({\bf k},\omega_n)$ with our $-\lambda I({\bf k},\omega_n)$, as it can be confirmed by considering their detailed expressions. Accordingly NSR calculations take into acount only the first term in the expansion Eq.(\ref{eqgpot1}), and uses the same approximate self-energy as we do.

\begin{figure}[htbp]
\begin{center}
\vbox to 60mm{\hspace{-30mm} \epsfysize=60mm \epsfbox{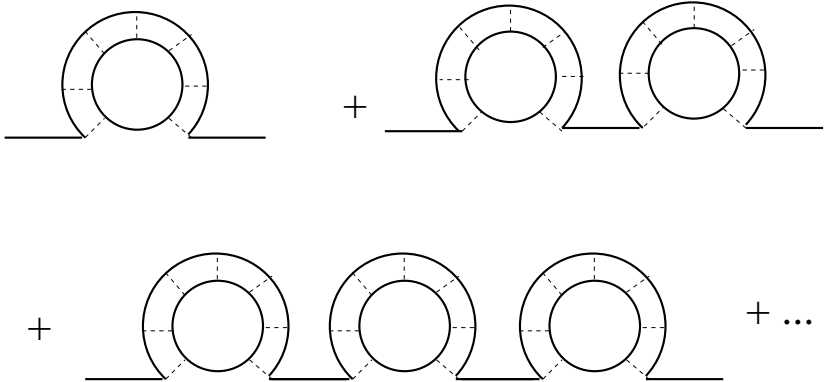}} 
\caption{Diagrams for the full propagator included in our approximation, in addition to the free particle propagator}
\label{diagrams}
\end{center}
\end{figure}

If one calculates directly the particle number by Eq.(\ref{eqn}), instead of using the thermodynamic potential $\Omega$  and $n=-\partial \Omega / \partial \mu $, the NSR scheme corresponds to take $G - G_0 = G_0 \Sigma G_0$, that is again the first order term in the above expansion of $G$, as it can be seen by performing the differentiation with respect to $\mu $ on Eq.(\ref{eqgpotnsr}).  By contrast our expression for $G$ in Eq.(\ref{eqn}) contains all orders in this expansion of $G$. Hence our scheme goes one step further by retaining all the terms in the expansion, while naturally keeping the same approximate expression for the self-energy. In other words we allow any number of self-energy inclusions in our propagator. We note that, in the dilute limit, it is perfectly valid to retain only the first order term, as NSR do. This is coherent with the fact that they find properly the zeroth order expression for the Bose condensation temperature. Diagrammatically they have for the propagator the familiar single "wheel" contribution (i.e. a ladder diagram with one side closed on itself). In our case we have any number of wheels interconnected as can be seen in Fig. \ref{diagrams}. Physically this means that, in  our case, the molecular bosons corresponding to the ladder diagrams can be broken and the elementary fermions present in the molecule can reappear individually.

\section{CRITICAL TEMPERATURES  AND  MOLECULAR  INSTABILITY}

In the case of the BCS transition temperature, for which $\mu >0$, the relation between the critical temperature $T_c$ and the chemical potential is given by the standard equation:
\begin{eqnarray}
- \frac{1}{\lambda}= \int_{0}^{\infty} dx \left[1\,-\,\frac{x^2}{x^2-1} \tanh \frac{x^2-1}{2t_c}\right]
\label{eqtcbcs}
\end{eqnarray}
where $t_c=T_c/\mu $. With the standard physical variables this equation reads:
\begin{eqnarray}
1=\frac{2a}{\pi}\int^{\infty}_{0}dk\;[1-\frac{\epsilon_k}{\epsilon_k -\mu}\tanh\frac{\epsilon_k -\mu}{2T_c}]
\label{eqn_k10}
\end{eqnarray}
where the chemical potential $\mu$ is positive. This equation is obtained by writing that $\Gamma ({\bf K},\omega)$ has a pole at the chemical potential for zero total momentum ${\bf K}=0$, that is by setting $Y_0=2$, $s=1$ and ${\bar K} \rightarrow 0$ in Eq. (\ref{eqY0}). This BCS transition line terminates when $\mu =0$ (this implies $t_c \rightarrow \infty$) which gives the relation between the limiting critical temperature $T_0$ and the corresponding scattering length $a_0$. This is more directly obtained from the equation with physical variables Eq.(\ref{eqn_k10}) and is given by $\pi /\sqrt{8ma_0^2T_0}= \int_{0}^{\infty}dz [1- \tanh(z^2/2)]$, i.e. explicitely $ma_{0}^{2}T_0=1.07=T_0/\epsilon _{b0}$ in terms of the molecular binding energy $\epsilon _{b0}=1/ma_0^2$ at this point. Obtaining $T_0/E_F$ and $1/k_Fa_0$ requires the numerical calculation of the gas density at that point (knowing that it satisfies $\mu =0$), along the procedure described in the preceding section \ref{mom}.

In the weak coupling limit $|a|k_F \ll 1$ this gives for the dominant order $T_c \sim E_F\exp(-1/\lambda)$. When the calculation is carried out to the next order term, one finds the numerical coefficient in front of the exponential as:
\begin{eqnarray}
T^{BCS}_c \simeq 0.61\, E_F\exp(-\frac{\pi}{2k_F|a|})
\label{eqn_k12}
\end{eqnarray}
However, already at the level of this coefficient, the simple weak coupling limit is not correct and one needs to take into account polarization contributions to the effective interaction, arising from second order diagrams in the gas parameter $\lambda$, as it has been shown by Gor'kov and Melik-Barkhudarov \cite{Gor'kov}. This leads to their well-known formula:
\begin{eqnarray}
T^{BCS}_c \simeq 0.28\,E_F\exp(-\frac{\pi}{2k_F|a|})
\label{eqn_k13}
\end{eqnarray}
Hence, at this level already, our approximation does not yield the exact result. In order to improve it on this side we should take into account polarization diagrams at the level considered by Gor'kov and Melik-Barkhudarov, and beyond as it has been done for example in Ref.\cite{kc}, \cite{Bar/Efr} and \cite{rctc}.

On the other hand we meet also naturally in our calculations the Bose-Einstein transition for the molecular states we have discussed in sections \ref{mol} and \ref{dilbec}. Indeed it is reached when the Bose factor $b(y_0/t)$ entering the self-energy in Eq. (\ref{eqsigm1}) diverges, since this factor corresponds physically to the statistical occupation of the molecular states. This divergence occurs for $y_0({\bar K})=0$. Naturally this is for a zero total molecular momentum ${\bar K}=0$ that it occurs first. This implies that $Y_0({\bar K})=y_0({\bar K})-{\bar K}^2/2+2s=-2$, since we need $s=-1$ in order to have ${\bar K}=0$ molecules, as noted above. From Eq. (\ref{eqY0}) we find for the critical temperature $T_c=|\mu |\,t_c$:
\begin{eqnarray}
\frac{1}{|\lambda|}= \int_{0}^{\infty} dx \left[1\,-\,\frac{x^2}{x^2+1} \tanh \frac{x^2+1}{2t_c}\right]
\label{eqtcbose}
\end{eqnarray}
where we use the fact that this transition occurs only for $\lambda<0$ to write $-\lambda=|\lambda|$. We see from this equation that, by letting $\mu \rightarrow 0$ (which implies again $t_c \rightarrow\infty$), the Bose-Einstein transition line terminates at the same point $\mu =0$ and $T=T_0$ as the BCS transition line. This Eq.(\ref{eqtcbose}) is actually in agreement with the finding of Ref. \cite{sdm} (as more easily seen when we go back to unreduced units). However, in contrast to the approach of these authors which treat the superfluid state, our reasoning involves only the normal state and its instabilities.

Going back to physical variables the above equation reads:
\begin{eqnarray}
1=\frac{2a}{\pi}\int^{\infty}_{0}dk\;[1-\frac{\epsilon_k}{\epsilon_k +|\mu |}\tanh\frac{\epsilon_k +|\mu |}{2T_c}]
\label{eqn_k14}
\end{eqnarray}
In the dilute molecular limit $k_F a \ll1$ (or equivalently $na^3 \ll 1$), we have $\epsilon _b \gg E_F$ and we have seen in section \ref{dilbec} that $\mu_B=2\mu+\epsilon _b$, implying $|\mu | > \epsilon _b /2$. This gives in Eq.(\ref{eqn_k14}) $|\mu | \gg T_c \sim E_F$ and hence
$\tanh [(\epsilon _k+|\mu|)/2T_c] \approx1$. Accordingly:
\begin{eqnarray}
1=\frac{2a}{\pi}\int^{\infty}_{0}dk\;\frac{|\mu|}{\epsilon_k +|\mu |}=\sqrt{2m|\mu | a^2}=\sqrt{2|\mu |/\epsilon _b }
\label{eqn_k15}
\end{eqnarray}
Hence in this limit our equation reduces to $\mu _B (T_c)=0$ as expected for the Bose-Einstein condensation of dilute gas. Using the standard formula for bosons \cite{LL} with
density $n_B=n$ and mass $2m$, one finds \cite{milstein}:
\begin{eqnarray}
T^{BEC}_c=3.31\,\frac{(n)^{2/3}}{2m} \simeq 0.218 E_F,
\label{eqn_k8}
\end{eqnarray}

Actually in this limit, our approximation is somewhat deficient as we have noted above in section
\ref{dilbec}. In fact in this regime $T \ll T_*$ we are dealing with a low-density bose-gas of
molecules with effective hard-core repulsion. The molecule-molecule
scattering length $a_{M}=0.6\,a$ has been first calculated recently by Petrov
\emph{et al} \cite{petrov} from the 4-particle Schr\"{o}dinger equation
and recently rederived diagrammatically by Brodsky \emph{et al} \cite{bkkcl}.
The inclusion of molecule-molecule interaction leads in this case
to a correction to critical temperature:
\begin{eqnarray}
T^{BEC}_c=0.218 \,E_F\;[1+\alpha(a_{M}n^{1/3})]
\label{eqn_k9}
\end{eqnarray}
where $\alpha \simeq 1.3$ is a numerical coefficient obtained by Kashurnikov, Prokof'ev,
and Svistunov \cite{Kashurnikov} in the framework of rather
complex Monte-Carlo calculations.

It is also worthwhile to note that actually, in the general case, we do not need the above physical interpretation of the Bose factor to find the Bose-Einstein transition. Indeed if we would calculate as discussed above the particle number as a function of temperature and chemical potential, and if we would enter the condensate domain of the phase diagram without realizing it, we would find that we could not accomodate all the particles we have in our system, even by further lowering the temperature, because the particle number would decrease with temperature. This would signal the appearance of a condensate in order to accomodate the remaining particles. In other words we can not miss the transition, which appears automatically in our approach.

Finally our major interest is the line in the phase diagram where molecules begin to form. As it has been stressed in Ref.\cite{rcmol} this threshold line is shifted from its standard location for molecules in vacuum, which is at unitarity. This is due to the presence of all the other fermions, which influence the formation of a molecule in much the same way as they do for the formation of Cooper pairs. Moreover, in contrast with the situation in vacuum and in similarity with Cooper pairs, the location of the threshold line where molecules begin to form depends on the total momentum of the molecule under consideration. For a very large momentum the molecule is insensitive to the presence of the other fermions, and the threshold line is at unitarity $a^{-1}=0$, just as in vacuum. On the other hand the molecules with zero momentum are the most sensitive to the presence of the other fermions. The question of molecular formation has already been considered in section \ref{mol} and the location of this threshold line is given by Eq.(\ref{eqlambseuil}). Again this line terminates at $\mu =0$, at the same point as the BCS line and the BEC line. Hence the three physical lines of interest in this section terminates at the same point.
\begin{figure}[htbp]
\begin{center}
\vbox to 80mm{ \epsfysize=80mm \epsfbox{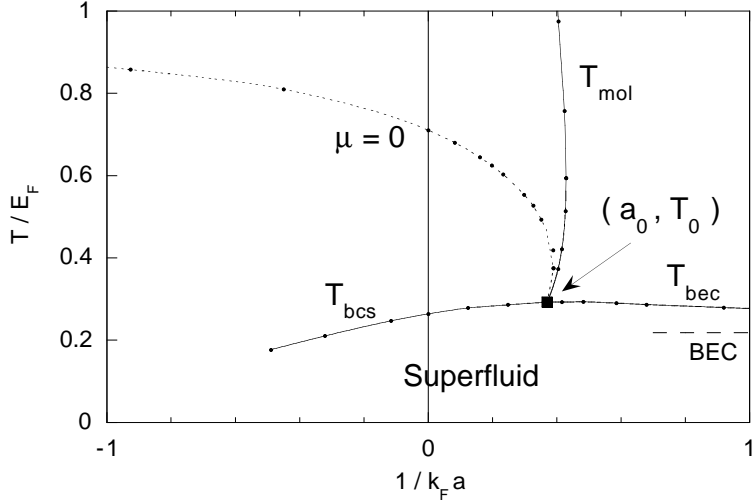}}
\caption{Threshold line $T_{mol}$ for the formation of molecules with zero total momentum ${\bf K}=0$, critical temperature for the BCS instability $T_{bcs}$ and critical temperature for the Bose-Einstein condensation of molecules $T_{bec}$, as a function of the scattering length $a$. Dots are the actual numerical results. Curves are smooth interpolations through these results. The wavevector $k_F$ is related to the gas density $n$ by $n=k_{F}^{3}/(6\pi ^2)$, and the Fermi energy is defined by $E_F= \hbar^{2}k_F^2/(2m)$. The line $\mu =0$ is drawn to show roughly the region (below this line) where the physics is qualitatively the one of a degenerate Fermi gas with $\mu >0$. The terminal point $a_0,T_0$ common to the molecular line, the BCS line and the BEC line is located at $1/(k_Fa_0)=0.37$ and $T_0/E_F=0.29$. The BEC limit for the dilute molecular gas is at $T_c = 0.218 E_F$}
\label{phasdiag}
\end{center}
\end{figure}
The last step in order to draw a physical phase diagram is to express the critical temperature in terms of the particle density $n$, rather than in terms of the chemical potential $\mu $. This is done by calculating numerically $n$ as a function of $\mu $, as indicated in section \ref{mom}. As indicated in the introduction, we use as usual wavevector and energy scales directly related to the atomic density $n$ for a single hyperfine state, namely the Fermi wavevector $k_F$ defined by $n=k_{F}^{3}/(6\pi ^2)$, and the Fermi energy $E_F= k_F^2/(2m)$. The results are displayed in Fig. \ref{phasdiag}. 

Let us first discuss the BCS and the BEC lines. Qualitatively the general shape is similar to the NSR result. The maximum of $T_c/E_F$ is very close to the terminal point $a_0,T_0$ and (at the scale of our figure) it is very smooth. This shape is also found in later work \cite{sdm,ppps}, including phenomenological approaches \cite{milstein}. Quantitatively our results are slightly higher than those of Perali \emph{et al} \cite{ppps} for the homogeneous case, as it can be seen from the $T_{c}^{h}$ in their Fig.1. Although this discrepancy is rather small, it is somewhat puzzling since both numerical calculations have been performed in a careful way, although the specific starting equations for the numerics are different as we have seen in section \ref{self}. Anyway the difference can be taken as a typical uncertainty arising in this kind of numerical calculations.

We consider finally the molecular threshold line. The quite interesting feature is that, for all the temperatures satisfying $T < E_F$ (and clearly even somewhat above), $1/k_F a$ stays typically around 0.4, that is roughly its value at the terminal point. This is in contrast with what one would expect from the fact that, at high temperature, this line goes toward the unitarity line $1/k_F a =0$ since this quantum shift of the molecular threshold clearly disappears in the classical regime. Hence the region between the molecular line and unitarity could have been pretty small, in contrast with what we find in Fig. \ref{phasdiag}. This makes naturally easier the experimental observation of the shift. It may even be that it has been already seen in the very recent vortex experiment \cite{vortex}. Indeed in order to observe vortices, experimentalists have to let the Fermi gas expand while at the same time forming molecules, in order to
see the depression in molecular density associated with the vortex in the same way as it is done for standard Bose condensates. In the experimental process, it has been observed that, when forming vortices on the BCS side $a<0$, it is necessary to switch the magnetic field to somewhere below the Feshbach resonance in order to be able to see the vortices (the corresponding $1/k_F a$ is of order 0.35, in nice agreement with the location of the molecular line). A possible explanation might be that, because (low momentum) molecules can not form beyond the molecular line, the formation of molecules necessary to see the vortices can not occur when the magnetic field is not brought to low enough values. However the answer lies in the trajectory of the Fermi gas in Fig.2 during expansion, which is not at all obvious to determine since the expansion is clearly a complicated dynamical process. Hence we can certainly not exclude that the explanation for this experimental observation lies somewhere else.

\section{CONCLUSION}

In this paper we have treated in a self-consistent way the various molecular instabilities arising in the normal state of a degenerate Fermi gas. We have stressed that these instabilities have to be contrasted with the smooth crossover in the nondegenerate
regime, corresponding to the temperature $T_*$ where the formation of the predominant molecules occurs as the temperature is lowered. We have covered all the range of scattering lengths, so as to cover the whole BEC-BCS crossover. The molecular instabilities manifest themselves mostly as poles of the vertex corresponding to particle-particle scattering. This vertex has been calculated in a ladder approximation. The threshold for the appearance of standard molecules corresponds to the existence of a pole at zero energy in the vertex. The BCS instability corresponds also to the appearance of molecular-like objects, namely Cooper pairs, but the pole in the vertex appears at the chemical potential. Finally the equation giving the Bose-Einstein condensation instability is a simple, but elegant, continuation of the equation giving the BCS instability. In order to find the critical temperatures corresponding to these various instabilities, we have taken into account the interactions between fermions in this normal gas, by making use of the same vertex as the one used to obtain the instabilities themselves. This leads to the self-consistent $T$-matrix approximation. We have shown that this approximation is quite satisfactory since it leads in a number of limiting cases to results which are in agreement with known exact results. These are specifically the high temperature regime, the large momentum limit at $T=0$, the dilute limit both on the BCS and the BEC side of the crossover, which we have investigated successively. The calculated phase diagram shows that, in particular, the threshold for formation of molecules at rest undergoes a sizeable shift toward the BEC side, due to the hindering effect of the quantum gas on the molecular wave function. This shift remains important up to temperatures comparable to the Fermi energy of the gas. Finally our approximation is somewhat defective on the BEC side since it does not allow to describe interaction between molecules. Improvement is clearly needed in this direction  and will be the subject of
separate publication. 

\section{ACKNOWLEDGEMENTS}

We are most grateful to T. Bourdel, Y. Castin, C. Cohen-Tannoudji, J. Dalibard, C. Salomon and A.M. Padokhin for stimulating discussions, and to Martin Zwierlein for discussions on the vortex experiment. Laboratoire de Physique Statistique is "associ\'e au Centre National
de la Recherche Scientifique et aux Universit\'es Paris 6 et Paris 7".
M.Yu.K. acknowledges the support of Russian Foundation for Basic
Research(Grant ¹ 04-02-16050), CRDF (Grant ¹ RP2-2355-MO-02) and
the grant of the Russian Ministry for Science and Education. He is
also grateful to the University Pierre and Marie Curie for the
hospitality on the first stage of this work.



\begin{references}
\bibitem{note}In the case $a>0$ we would in addition a pole for $\Sigma(\omega )$, corresponding to the presence of a bound state (see section VI). This would give rise to an additional term when the contour is deformed, leading to the bound state contribution in the Beth-Uhlenbeck result \cite{beuh}.
\end{references}
\end{document}